\documentclass{article}

\usepackage[utf8]{inputenc} 
\usepackage[T1]{fontenc}    
\usepackage{hyperref}       
\usepackage{url}            
\usepackage{booktabs}       
\usepackage{amsfonts}       
\usepackage{nicefrac}       
\usepackage{microtype}      
\usepackage{graphicx}
\usepackage[preprint]{neurips_2019}

\usepackage{natbib}
\title{SAT-based Compressive Sensing}

\author{%
  Ramin Ayanzadeh\\
  Department of Computer Science and Electrical Engineering\\
  University of Maryland, Baltimore County\\
  Baltimore, MD 21250 \\
  \texttt{ayanzadeh@umbc.edu} \\
     \And
   Milton Halem \\
   Department of Computer Science and Electrical Engineering\\
   University of Maryland, Baltimore County\\
   Baltimore, MD 21250 \\
   \texttt{halem@umbc.edu} \\
   \AND
   Tim Finin \\
   Department of Computer Science and Electrical Engineering\\
   University of Maryland, Baltimore County\\
   Baltimore, MD 21250 \\
   \texttt{finin@umbc.edu} \\
  }

\begin{document}

\maketitle

\begin{abstract}
We propose to reduce the original well-posed problem of compressive sensing to weighted-MAX-SAT. Compressive sensing is a novel randomized data acquisition approach that linearly samples sparse or compressible signals at a rate much below the Nyquist-Shannon sampling rate. The original problem of compressive sensing in sparse recovery is NP-hard; therefore, in addition to restrictions for the uniqueness of the sparse solution, the coding matrix has also to satisfy additional stringent constraints—usually the restricted isometry property (RIP)— so we can handle it by its convex or nonconvex relaxations. In practice, such constraints are not only intractable to be verified but also invalid in broad applications. This paper bridges the gap between employing modern SAT solvers and a vast variety of compressive sensing based real-world applications. We first divide the well-posed problem of compressive sensing into relaxed sub-problems and represent them as separate SAT instances in conjunctive normal form (CNF). After merging the resulting sub-problems, we assign weights to all clauses in such a way that the aggregated weighted-MAX-SAT can guarantee successful recovery of the original signal. The only requirement in our approach is the solution uniqueness of the associated problems, which is notably looser. As a proof of concept, we demonstrate the applicability of our approach in tackling the original problem of binary compressive sensing with binary design matrices. Experimental results demonstrate the supremacy of the proposed SAT-based compressive sensing over the $\ell_1$-minimization in the robust recovery of sparse binary signals. SAT-based compressive sensing on average requires 8.3\% fewer measurements for exact recovery of highly sparse binary signals ($s/N\approx 0.1$). When $s/N \approx 0.5$, the $\ell_1$-minimization on average requires 22.2\% more measurements for exact reconstruction of the binary signals. Thus, the proposed SAT-based compressive sensing is less sensitive to the sparsity of the original signals.  
\end{abstract}

\section{Introduction}
Compressive sensing (also known as compressed sensing, compressive sampling or sparse sampling) is a randomized data acquisition method that linearly samples sparse or compressible signals at a rate much below of the Nyquist-Shannon sampling theorem (\cite{candes2004robust, donoho2006compressed, candes2004near}). Compressive sensing outperforms traditional data acquisition approaches where: (a) traditional sensing techniques are very time-consuming –i.e., MRI and functional MRI; (b) energy efficiency is vital –e.g., wireless sensor networks (WSN) and wireless body sensor nodes (WBSN); (c) sensing is too expensive (namely high-speed ADCs); (d) we have to utilize few sensors like non-visible wavelengths (\cite{rani2018systematic}). From a data acquisition point of view, instead of sensing $N$ samples uniformly and then compressing them into a vector of size ${s},$ compressive sensing performs ${m}$ linear measurements such that ${m}$ is reasonably close to ${s}.$ In other words, compressive sensing performs both sensing and size reduction tasks simultaneously. Since the majority of signals are either self-sparse in their original domain or have a sparse representation in some transform domain, compressive sensing samples the signal at a rate much below the Nyquist sampling rate –in many real-world applications $m=O\left(s\ln(N/s)\right)$ (\cite{foucart2013mathematical}). 

In traditional signal processing approaches, sensing is much more complicated than the recovery of the original signals. In compressive sensing, however, the encoding process is simple, and the signal reconstruction is NP-hard. For a given measurement vector $y\in {{\mathbb{R}}^{m}}$ and a design matrix $A\in {{\mathbb{R}}^{m\times N}}$ with $m\ll N$, the original problem of compressive sensing aims to recover a sparse signal $x\in {{\mathbb{R}}^{N}}$ such that $y=Ax$. This is an underdetermined system which generally has infinite solutions. Therefore, compressive sensing imposes some constraints on the coding matrix–usually the restricted isometry property (RIP)— to guarantee that the sparse solution is unique. Let $\|x{{\|}_{0}}$ stand for the sparsity level of $x$, i.e., the number of nonzero entries of $x$,  the ultimate goal of compressive sensing  can be formulated as follows:
\begin{equation}
    \label{eqn:ell_0}
    {{\min }_{x\in {{\mathbb{R}}^{\mathbb{N}}}}}\|x{{\|}_{0}}\quad \textnormal{subject to}\quad Ax=y,
\end{equation}

where we exploit the sparsity of ${x}$ through $\ell_0$-minimization to reconstruct the original signal from far fewer samples than required by the sampling theorem (\cite{foucart2013mathematical}). From a complexity perspective, finding the minimal support set of ${x}$ in problem (\ref{eqn:ell_0}) is NP-hard (\cite{muthukrishnan2005data}). Hence, well-posed compressive sensing is intractable in the realm of classical computing (\cite{ayanzadeh2019quantum}). Greedy algorithms like orthogonal matching pursuit (OMP) can tackle the $\ell_0$-minimization under some restrictive constraints (\cite{foucart2013mathematical}). In many real-world applications, however, greedy algorithms suffer in recovering signals with high-enough accuracy (\cite{eldar2012compressed}). From an application viewpoint, compressive sensing started to revolutionize the real-world applications through convexifying the problem (1). Since $\ell_p$-norm approaches $\ell_0$-norm when nonnegative $p\downarrow 0,$ we can represent the problem (1) as follows:
    \begin{equation}
        \label{eqn:ell_p}
        {{\min }_{x\in {{\mathbb{R}}^{\mathbb{N}}}}}\|x{{\|}_{p}}\quad \textnormal{subject to}\quad Ax=y.
    \end{equation}

Here, we need to provide necessary and sufficient conditions to guarantee that both problems (\ref{eqn:ell_0}) and (\ref{eqn:ell_p}) appoint an identical solution. On this basis, we generally apply additional stringent constraints on the design matrix to certify this possibility using tools such as restricted isometry constants or null and range space properties (\cite{baraniuk2008simple}).

For $p>1,$ the unique solution of this strictly convex problem is generically full-support, i.e., each of its components is non-zero (\cite{shen2018least}).  We can reduce the problem (\ref{eqn:ell_p}) to a linear program when $p=1,$ and necessary and sufficient conditions are available to guarantee the uniqueness of the optimal solution (\cite{mousavi2017solution}). Convex optimization based methods like basis pursuit (BP), Dantzig selector, and gradient-based algorithms generally require significantly more computational resources, but they can outperform other ill-posed techniques like greedy algorithms and hybrid approaches (i.e., compressive sampling matching pursuit and stage-wise OMP) in terms of recovery accuracy (\cite{foucart2013mathematical}). The case of $p \in (0,1)$ leads to a nonconvex objective function, although it obtains not only much less restrictive conditions on the design matrix but also more robust and stable theoretical guarantees at the cost of higher complexity in the recovery phase (\cite{chartrand2009fast}). As an illustration, a sufficient condition for successful recovery in noiseless environments through $\ell_{0.5}$-norm is significantly less restrictive than the analogous results for $\ell_1$-norm recovery (\cite{saab2008stable}). Recent studies have demonstrated that the well-posed problem of compressive sensing is tractable by adiabatic quantum computers (\cite{ayanzadeh2019quantum}); however, the proposed approach is limited to only recover the sparse binary signals. 

This paper proposes to tackle the well-posed problem of compressive sensing (for $p=0$) via reducing the $\ell_0$-minimization—shown in (\ref{eqn:ell_0})— to Weighted-MAX-SAT instances. The Boolean satisfiability (SAT) is the problem of determining whether a given Boolean formula can be interpreted as “True” with a constant replacement of values (“True” or “False”) for all Boolean variables (\cite{cormen2009introduction}). This problem is NP-complete (\cite{cook1971complexity}); however, various real-world applications of SAT have revealed that worst cases are less likely to happen in practice. Hence, the modern SAT solvers can tackle the SAT instances with thousands of variables and millions of clauses (in clausal normal form). We first define two ill-posed problems via relaxation of constraints in (\ref{eqn:ell_0}) and represent them as independent SAT instances –in clausal normal form (CNF)— over the same variables. After merging these two SAT instances, we assign weights to all clauses and provide sufficient conditions to guarantee that the resulting Weighted-MAX-SAT and the original problem of compressive sensing will appoint an identical solution. As a proof of concept, we used the Z3 framework to implement the proposed method for recovery of binary signals with binary coding matrices. Experimental results demonstrated the supremacy of the proposed SAT-based compressive sensing over the $\ell_1$-minimization in the robust recovery of sparse binary signals. 

\section{Boolean Satisfiability}
The problem of Boolean satisfiability (also known as propositional satisfiability, satisfiability or SAT) aims to determine whether a given Boolean expression/formula can be interpreted as “True” with a constant replacement of the values (“True” or “False”) for all Boolean variables (\cite{cormen2009introduction}). A Boolean formula is in conjunctive/clausal normal form (CNF) if it is a conjunction of clauses, where each clause is the disjunction of literals –Boolean variables or their negation (Gu 1994). When we restrict the clauses to contain at most $k$ literals, for  $k\ge 3,$ the resulting $k$-SAT instances are NP-complete (\cite{calabro2006duality, impagliazzo2001complexity}). We can convert any propositional expression to CNF in polynomial-time, and the majority of the modern SAT solvers have adapted it for standardizing the representation of the SAT instances (\cite{gu1999algorithms,biere2009handbook}). 

Satisfiability is the first NP-complete problem (\cite{cook1971complexity}). According to Cook-Levin theorem, we can reduce all problems in the NP class to the SAT instances in polynomial time (\cite{cormen2009introduction}). Real-world problems, however, have complicated properties that make the use of the standard SAT solvers challenging. Therefore, we extend the search-oriented nature of the original SAT to cover other problem types–including but not limited to optimization and model counting (\cite{biere2009handbook,marques2000boolean}). For a given Boolean formula in CNF, the MAX-SAT (maximum satisfiability) determines the maximum number of satisfiable clauses. From a problem formulation attitude, the MAX-SAT adds optimization perspective to the original SAT, which has searching nature. Weighted-MAX-SAT is an extended version of the MAX-SAT in which, each clause has an associated weight that specifies the penalty of not satisfying the clauses.

Solving satisfiability instances can require exponentially large computational resources. Worst cases, however, are less likely to happen in practice and modern SAT solvers can handle the real-world applications with thousands of variables and millions of clauses. As an illustration, advances in developing efficient heuristics have resulted in commercialized tools for SAT-based planning applications (\cite{kautz2006satplan04}). Since we can transfer first-order logic formulas into propositional logic formulas (\cite{russell2016artificial}), SAT serves as the cornerstone for a vast range of AI applications –more precisely, reasoning and inference systems. For example, “Z3” is an efficient satisfiability modulo theory (SMT) solver that can satisfy higher order expressions (\cite{de2008z3}). 

Furthermore, SAT solvers have demonstrated remarkable functionalities in a vast variety of electronic design automation (EDA) applications–including but not limited to combinational equivalence checking, logic optimization, functional test vector generation, test pattern generation, circuit delay computation and bounded model checking (\cite{marques2000boolean}). Satisfiability has also started to address various problems in cryptography. In 2005, for instance, high-performance SAT solvers were able to break several standard cryptographic hash functions (\cite{mironov2006applications}). In the same way, parallel SAT solvers have shown to be applicable for integer factoring applications (\cite{lunden2015factoring}).

\section{SAT-based Compressive Sensing}
From a problem-solving perspective, the recovery module in compressive sensing receives the measurement vector $y\in {{\mathbb{R}}^{m}}$ and the design matrix $A\in {{\mathbb{R}}^{m\times N}}$ as input, and recovers a sparse vector $x\in {{\mathbb{R}}^{N}}$ that: (a) ${x}$ satisfies $y=Ax;$ and (b) ${x}$ has maximum number of zeros. In SAT-based compressive sensing, we first define two sub-problems via relaxing the key aspects of the original problem of compressive sensing –shown in (\ref{eqn:ell_0}). We define the first sub-problem as $f_1 := y=Ax$ Where we have neglected the sparsity constraint of (\ref{eqn:ell_0}).  

We define the second sub-problem as $f_2 := {argmin}_x {\| {x}\|_0}$ where the global optimum appears when the zero vector  $x=0.$ Afterward, we represent $f_1$ and $f_2$ as two separate SAT instances (but over the same variable ${x}$) in CNF  as follow:

\begin{equation}
    \label{eqn:f_1}
    {{f}_{1}}:=C_{1}^{1}\wedge C_{1}^{2}\wedge \cdots \wedge C_{1}^{p_1};
    \end{equation}

\begin{equation}
    \label{eqn:f_2}
    {{f}_{2}}:=C_{2}^{1}\wedge C_{2}^{2}\wedge \cdots \wedge C_{2}^{p_2},
    \end{equation}

    where $C_{1}^{i}$ for $i \in \{1,2,\dots,p_1\}$ and $C_2^j$ for $j\in \{1,2,\ldots ,{{p}_{2}}\}$ are clauses over the elements of the vector ${x}$ in $f_1$ and $f_2$ respectively. Since both problems (\ref{eqn:f_1}) and (\ref{eqn:f_2}) are in CNF, $f := f_1 \wedge f_2$ will be a Boolean expression in CNF—shown in (\ref{eqn:f}). Hence, any ${x}$ that satisfies (\ref{eqn:f}) is guaranteed to satisfy (\ref{eqn:f_1}) and (\ref{eqn:f_2}) simultaneously. 

    \begin{equation}
        \label{eqn:f}
        {f} := {{f}_{1}} \wedge {{f}_{2}} = C_{1}^{1}\wedge C_{1}^{2}\wedge \cdots \wedge C_{1}^{p_1} \wedge C_{2}^{1}\wedge C_{2}^{2}\wedge \cdots \wedge C_{2}^{p_2}.
        \end{equation}

        Finally, we assign weights to clauses in (\ref{eqn:f}) to form a Weighted-MAX-SAT problem as follow: 

\begin{equation}
    \label{eqn:cs_sat}
    f_{CS} := {w_1}{C_1^1} \wedge {w_1}{C_1^2} \wedge \dots \wedge  {w_1}{C_1^{p_1}} \wedge {w_2}{C_2^1} \wedge {w_2}{C_2^2} \wedge \dots \wedge {w_2}{C_2^{p_2}}.
    \end{equation}

    We only need to provide necessary and sufficient conditions to guarantee that both problems (\ref{eqn:ell_0}) and (\ref{eqn:cs_sat}) appoint an identical solution. Assuming that the sparse solution of (\ref{eqn:ell_0}) is unique (under appropriate conditions), satisfying (\ref{eqn:w_condition}) guarantees that (\ref{eqn:cs_sat}) has a unique global optimum which is identical to the sparse solution of (\ref{eqn:ell_0}).

    \begin{equation}
        \label{eqn:w_condition}
        w_1 > \sum_{j=1}^{p_2}{w_2^j}.\quad\quad \textnormal{for} \ i=1,2,\dots,{p_1}.
        \end{equation}

It is crucial to highlight that we do not solve $f_1$ and $f_2$ independently. Indeed, we solve (8) which is the SAT-based representation of the original problem of compressive sensing. In this formulation, $f_1$ defines the feasible domain for $f_2$, and $f_2$ exploits the feasible space to minimize the $\| {x}\|_0.$

\section{Proof of Concept }
Restricting the vector ${x}$ in the problem (\ref{eqn:ell_0}) to take its values from $\{0,1\}$ leads to an NP-hard discrete optimization problem –called binary compressive sensing (\cite{nakarmi2012bcs,liu2011formulating}). From an application point of view, not only binary signal sources have real-world applications (i.e., event detection in wireless sensor networks, group testing, spectrum hole detection for cognitive radios, etc.), but we also can leverage it to other types of signals (\cite{nakarmi2012bcs}). As a proof of concept, therefore, we demonstrate that we can employ the proposed SAT-based compressive sensing approach to tackle the well-posed problem of binary compressive sensing.

While most of the studies in compressive sensing focus on random design matrices with Gaussian distribution, solid theoretical and experimental results are available to guarantee that we can also exactly recover sparse (or compressible) signals from random binary matrices with a very high probability (\cite{zhang2010compressed}). In the realm of binary compressive sensing, however, binary coding matrices have demonstrated remarkably lower performance compared to non-binary design matrices (\cite{nakarmi2012bcs,liu2011formulating}). As an illustration, sparse random design matrices with Bernoulli distribution can only recover highly sparse binary signals –${s/N}<0.1$ (Ref{BCS}). Thus, several studies have focused on generating non-binary design matrices to remediate the performance and quality of binary compressive sensing techniques (\cite{nakarmi2012bcs}). Therefore, without losing the generality of the proposed SAT-based compressive sensing, we focus on the most challenging arrangement where both ${x}$ and ${A}$ are binary.  

For a given measurement vector $y \in \mathbb{N}^m$ and a coding matrix $A \in \{0,1\}^{m\times  N}$ (where $m\ll N$), the objective in SAT-based binary compressive sensing is to construct a Weighted-MAX-SAT instance, shown in (\ref{eqn:cs_sat}), over the binary vector $x \in \{0,1\}^N$ and guarantee that (\ref{eqn:cs_sat}) and (\ref{eqn:ell_0}) appoint an identical solution. We start with defining the relaxed sub-problem ${{f}_{1}},$ shown in (\ref{eqn:f_1}), which tries to only solve $y=Ax.$ Because both ${A}$ and ${x}$ take their values from $\{0,1\},$ calculating $\langle A_{i.},x \rangle$ is equivalent to finding the Hamming weight (or population count) of $\{{{A}_{i1}}{{x}_{1}},{{A}_{i2}}{{x}_{2}},\ldots ,{{A}_{iN}}{{x}_{N}}\}$ as follows:

\[
    {{z}_{i}}=\langle {{A}_{i.}},x\rangle =\sum\limits_{j=1}^{N}{{{A}_{ij}}{{x}_{j}}};
\]

and represent the relaxed sub-problem $f_1$ as $y_i = z_i, \quad \forall i \in [1,m].$ Since both ${y}$ and ${z}$ take their values from ${[0,N]},$ we can represent them in binary basis as $\hat{y}\in {{\{0,1\}}^{m\left( \left\lfloor {{\log }_{2}}N \right\rfloor +1 \right)}},$and $\hat{z}\in {{\{0,1\}}^{m\left( \left\lfloor {{\log }_{2}}N \right\rfloor +1 \right)}}$ respectively. Afterward, we can use the “XNOR” Boolean operator for representing the $f_1$ in binary basis as follows:

\[
    {{f}_{1}}:=\neg ({{\hat{y}}_{i}}\oplus {{\hat{z}}_{i}}),\quad \forall i\in \left[ 1,m\left( \left\lfloor {{\log }_{2}}N \right\rfloor +1 \right) \right].
    \]

Finally, we can convert the Boolean expressions $\neg{(\hat{y}_i \oplus \hat{z}_i)}$ to CNF independently, and merge the resulting sub-SAT instances via the “AND” operator to represent $f_1 : \quad y=Ax$ as an SAT in CNF. For the given measurement vector ${y},$ we can convert the basis and form the binary vector $\hat{y}.$ To construct the binary vector $\hat{z},$ we can employ the gate model of “half-adder” and “full-adder” modules in a tree-based structure and build a digital circuit for the problem of Hamming weight. 

\begin{figure}  
    \centering
    \includegraphics[scale=1]{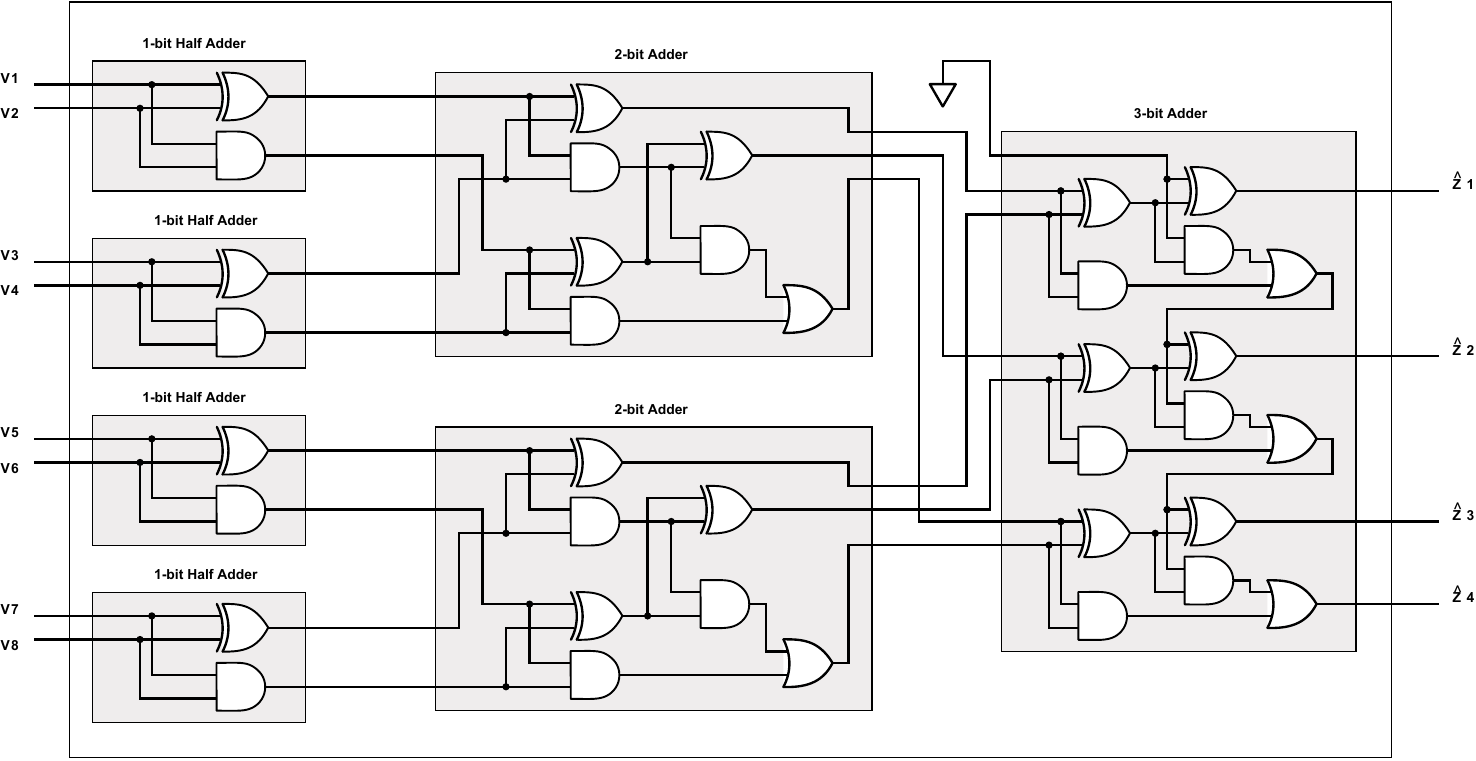}
    \caption{Gate model of a Hamming weight module for an 8-bit input vector $v$}
    \label{fig:hamming}
  \end{figure}

\begin{figure}
    \centering
    \includegraphics[scale=1]{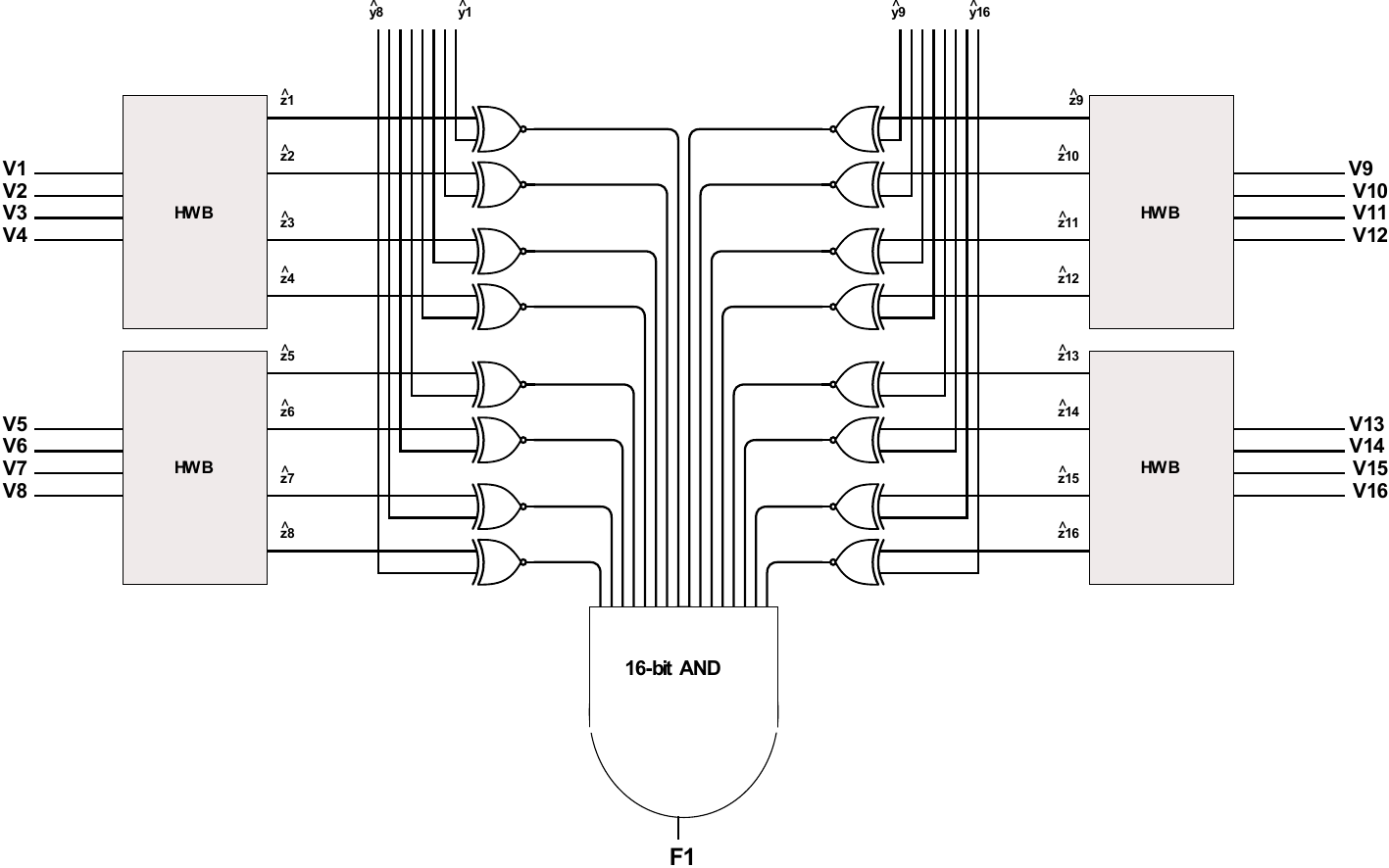}
    \caption{Gate model of the relaxed sub-problem $f_1 : y=Ax$ for an 8-bit input vector ${x}$}
    \label{fig:inner_product}
  \end{figure}

Figure \ref{fig:hamming} illustrates an eight-bit Hamming weight module to generate the result of $\langle A_{i.},x \rangle.$ In this tree-based structure, the Hamming weight module will require $O\left({\log_2{{p_B}{N}}}\right)$ full-adders where $p_B$ represents the Bernoulli distribution parameter. Hence, the encoding process can also take the advantages of sparse coding matrices which generally result in smaller Boolean expressions. Figure \ref{fig:inner_product} illustrates a prototype circuit for constructing the Boolean expression for representing $f_1$ when $N=8$ and $m=4.$

We know that the global minimum of $\| {x}\|_0$ occurs at the zero vector $x=0.$ Therefore, we can represent the second sub-problem $f_2$ in CNF as follows:
\[
    f_2 := \neg{x_1} \wedge \neg{x_2} \wedge \dots \wedge \neg{x_N}.
    \]
Finally, we need to satisfy (\ref{eqn:w_condition}) to guarantee that (\ref{eqn:ell_0}) and (\ref{eqn:cs_sat}) ( will appoint an identical solution. Because ${x}$ is binary, we can represent (9) as below: 
\[w_1 > Nw_2.\]

\section{Experimental Results }
For implementing the proposed symbolic SAT-based compressive sensing, we used the Z3 which is a theorem prover by Microsoft Research. Since our framework performs symbolic computing, we simplify the output of each implemented module (i.e., full-adder, Hamming weight, and inner-product modules) to subside the bloating of the Boolean expressions in successive calls of modules. Our empirical studies demonstrated that simplifying the Boolean equations constantly improves the performance of the SAT solvers, albeit more modeling time. To satisfy/optimize the ultimate Weighted-MAX-SAT, which represents the well-posed problem of compressive sensing, we used the core-guided solver with compressed MAX-SAT resolution (\cite{narodytska2014maximum}) which is also available in Z3. In the following experiments, we have compared the proposed SAT-based compressive sensing with $\ell_1$-norm recovery.

In our first experiment, we measure the oversampling factor ($m/{s \log{N/s}}$) for different sparsity rates ($s/N \in \left(0, 1 \right]$. To this end, for a fixed signal size (here, $N=30$) and a specific sparsity rate, we generated 10 independent random examples. For each test instance, we generate the coding matrix randomly with Bernoulli distribution ($p_B=0.5$). After generating the signal randomly (with the specified sparsity rate), we construct the measurement vector through $y=Ax$. For each method, the objective in this experiment was to find an appropriate ${m}$ such that we can exactly recover all 10 randomly generated test instances. In other words, instead of measuring the average recovery error, we adjust the number of required measurements to recover all elements of ${x}$ correctly. 

\begin{figure}  
  \centering
  \includegraphics[scale=0.9]{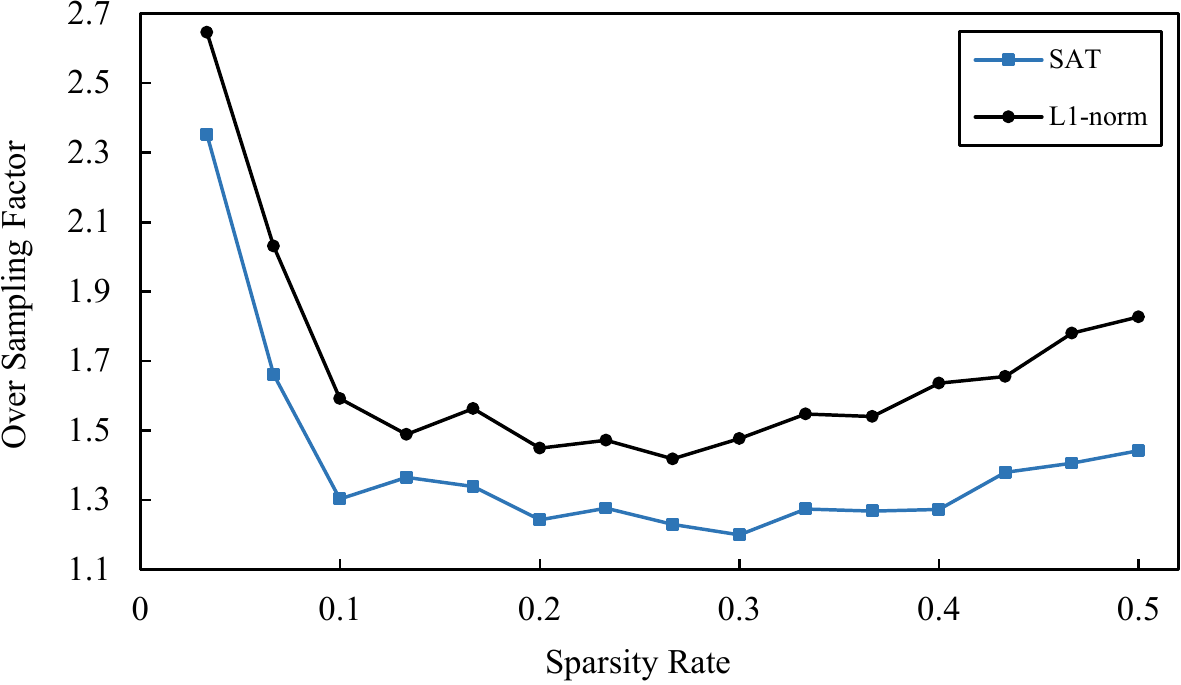}
  \caption{Performance comparison between SAT-based and $\ell_1$-minimization recovery based on optimum oversampling factor ($m/{s \log{N/s}}$) for different sparsity rate ($s/N \in \left(0,0.5\right]$)}
  \label{fig:test_1}
\end{figure}

Figure \ref{fig:test_1} illustrates the results of this experiment, which reveals the supremacy of the proposed SAT-based compressive sensing over the $\ell_1$-norm recovery. For highly sparse signals ($s/N \approx 0.1$), the $\ell_1$-norm recovery needs on average 8.3\% more measurements than the SAT-based compressive sensing. Figure \ref{fig:test_1} also shows that the SAT-based compressive sensing is less sensitive to the sparsity rate. More precisely, when the sparsity rate increases, the distance between the oversampling factors of $\ell_1$-norm recovery and SAT-based compressive sensing increases. As an illustration, for $s/N=0.5$, the SAT-based compressive sensing can exactly recover the signals with 22.2\% fewer measurements.

\begin{figure}[!h]  
  \centering    
  \includegraphics[scale=0.9]{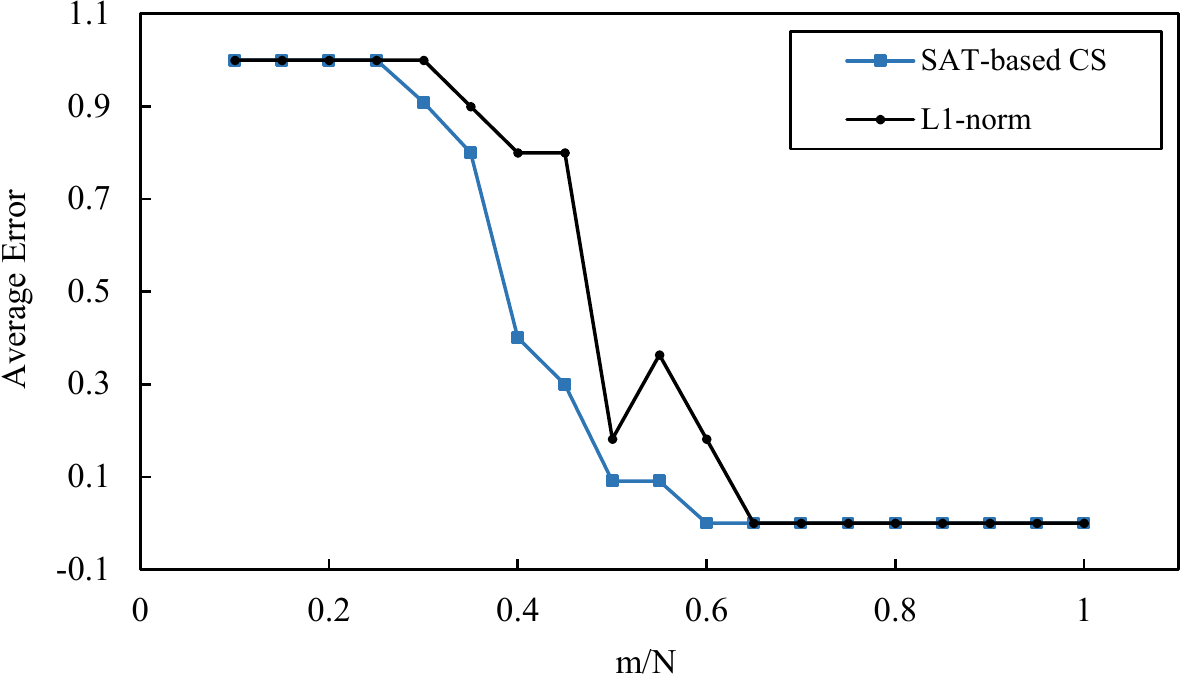}
  \caption{Recovery performance comparison  for $p_B=0.5$ and $s/N=0.5$}
  \label{fig:test_2_1}
\end{figure}

\begin{figure}[!htp]  
  \centering    
  \includegraphics[scale=0.9]{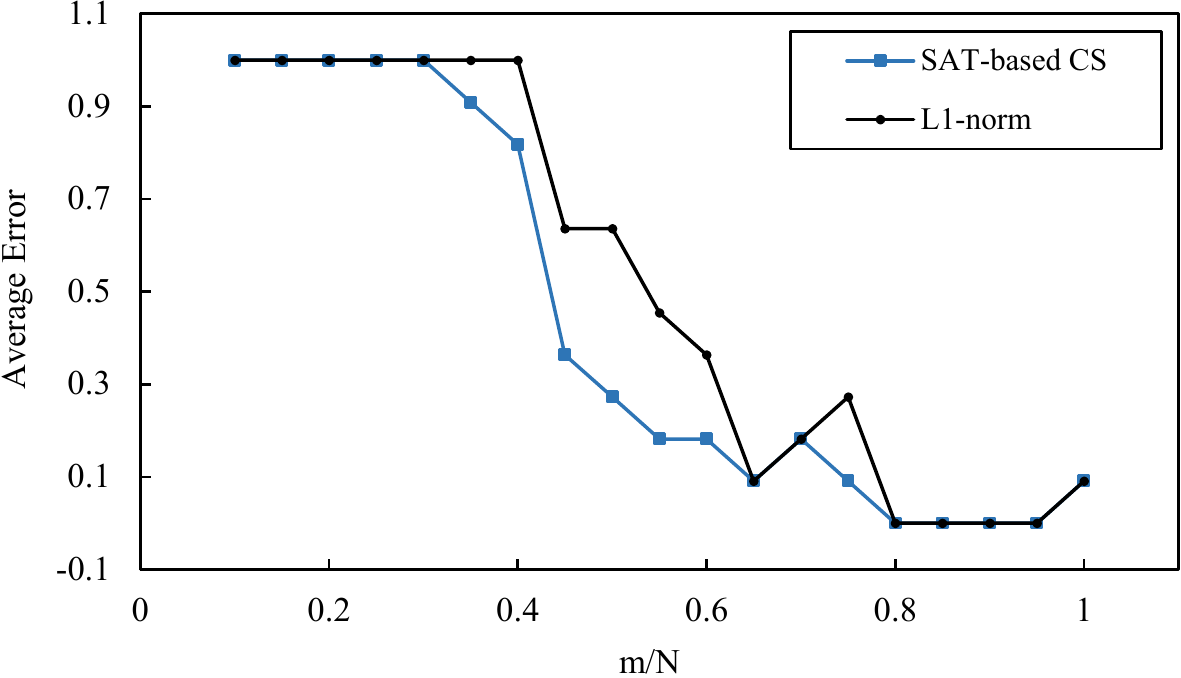}
  \caption{Recovery performance comparison for $p_B=0.3$ and $s/N=0.3$}
  \label{fig:test_2_2}
\end{figure}

In the second experiment, we keep the sparsity rate fixed and measure the average recovery error for different compression rates ($m/N \in \left[0.1, 1\right]$). To this end, for each specific compression rate, we generate 10 random test instances (for different signal size $N \in \left[20, 30 \right]$) and measure the average recovery error over all test instances. Since $x \in \{0,1\}$, we divide the Hamming distance between the recovered and original signals by ${N}$ for representing the error for each recovery. Figure \ref{fig:test_2_1} illustrates the experiment results for $p_B=0.5$ and $s/N = 0.5$. Similarly, Figure \ref{fig:test_2_2} shows the experiment results for $p_B=0.3$ and $s/N=0.3$. Not only the SAT-based compressive sensing provides higher performance, but it also demonstrates more robust functionality. 

\section{Discussion}
Compressive sensing is a randomized data acquisition approach that linearly samples sparse or compressible signals at a rate much below the Nyquist-Shannon rate. The well-posed problem of compressive sensing is intractable in the realm of classical computing. Therefore, we impose additional stringent constraints on the coding matrix (usually RIP) and tackle the ill-posed version of it. Such constraints are not only NP-hard to be verified but also invalid in some applications. In this paper, we proposed to reduce the original problem of compressive sensing to the weighted-MAX-SAT and tackle the well-posed problem of compressive sensing via modern SAT solvers. We first divide the original well-posed problem of compressive sensing into two relaxed sub-problems and represent them as separate SAT instances in CNF. In the first sub-problem, we relax the sparsity constraint and formulate the resulting sub-problem only to solve the given undetermined system. The second sub-problem aims only to find the sparsest signal, which we know where the global optimum appears in advance. Afterward, we merge the resulting CNF instances via the “AND” operator and form a larger SAT in CNF. Finally, we assign weights to the clauses of the aggregated CNF in such a way that the ultimate weighted-MAX-SAT and the original problem of interest are guaranteed to appoint an identical solution. 

Remark that the only required assumption here is to have a unique solution for the original problem of compressive sensing, which is much looser than those of relaxations or greedy methods. As a proof of concept, we demonstrated the applicability of our method in tackling the well-posed problem of binary compressive sensing with binary design matrices. Experimental results revealed that the proposed SAT-based compressive sensing outperforms the $\ell_1$-norm based recovery in terms of not only the oversampling factor but also average recovery error. Also, experimental results demonstrated that the proposed SAT-based compressive sensing is less sensitive to the sparsity rate. More precisely, when the sparsity rate increases, the distance between the oversampling factors also increases. 

To implement the symbolic computations, we used the Z3 framework, which is a theorem prover by Microsoft Research. Symbolic computing generally requires more computational resources, so we need to utilize high-performance computing for large-scale SAT-based problem-solving. The current release of Z3, however, does not support distributed computing for MAX-SAT. Thus, we performed experiments with remarkably small signals. Although the proposed SAT-based compressive sensing was able to outperform the $\ell_1$-norm recovery technique, it requires more computational resources. Hence, in practice, SAT-based compressive sensing is a proper choice where the lower compression rate ($m/N.$) is crucial. 

This paper bridges the gap between the trends in advancing the SAT solvers and a broad range of compressive sensing based real-world applications. In standard compressive sensing, we assume that the measurements come from noiseless sources, and we also have perfect knowledge about the coding matrix. In practice, however, such assumptions are not valid. As future work, we are extending the proposed SAT-based compressive sensing to handle the noisy measurements. Besides, we are leveraging the proposed model for tackling the problem of compressive sensing with matrix uncertainty, which is a more general problem that handles the case where we only can approximate the design matrices. It is worth noting that we can leverage the proposed approach for different models of quantum computers (like gate and adiabatic models).

\subsubsection*{Acknowledgments}

This research has been supported by NASA grant (\#NNH16ZDA001N-AIST 16-0091), NIH-NIGMS Initiative for Maximizing Student Development Grant (2 R25-GM55036), and the Google Lime scholarship. We would like to thank the NSF fund-ed Center for Hybrid Multicore Productivity Research for access support to IBM Min-sky cluster. We would also like to thank the D-Wave Systems management team, namely Rene Copeland, for granting access to the D-Wave 2000 quantum annealer.

\bibliography{references}

\end{document}